\documentstyle[twocolumn,pre,aps]{revtex}

\begin{document}

\relax 

\draft \preprint{RRU/3-cg} 

\title{Note on Two Theorems in Nonequilibrium Statistical Mechanics}

\author{E. G. D. Cohen${}^1$, G. Gallavotti${}^2$}
\address{ ${}^1$The Rockefeller University, New York, NY 10021,
USA\\ 
${}^2$Fisica, Universit\'{a} di Roma, ``La Sapienza'', 00185
Roma, Italia}

\date{\today}
\maketitle
\begin{abstract}

An attempt is made to clarify the difference between a theorem derived
by Evans and Searles in 1994 on the statistics of trajectories in phase
space and a theorem proved by the authors in 1995 on the statistics of
f\/luctuations on phase space trajectory segments in a nonequilibrium
stationary state.

\end{abstract}

\narrowtext

Recently a Fluctuation Theorem (FT) has been proved by the authors
(GC), \cite{[GC95]}, for f\/luctuations in nonequilibrium stationary
states.  Considerable confusion has been generated about the
connection of this theorem and an earlier one by Evans and Searles
(ES), \cite{[ES94]}, so that it seemed worthwhile to try to clarify
the present situation with regards to these two theorems.

In a paper in 1993 by Evans, Cohen and Morriss \cite{[ECM93]}, theoretical
considerations lead them to a computer experiment about the statistical
properties of the f\/luctuations of a shear stress model (viscous current) -
or the related entropy production rate - in a thermostatted
sheared viscous f\/luid (plane Couette f\/low) in a nonequilibrium
stationary state.

The Fluctuation Relation found in the simulation \cite{[ECM93]}, reads
in current notation, \cite{[GC95]}:
\begin{equation}
\frac{\pi_\tau(p)}{\pi_\tau(-p)} \simeq e^{\tau \sigma_+p}
\label{1}
\end{equation}
Here $\pi_\tau(p)$ is the probability of observing an average phase
space contraction rate (which in the models considered has the
interpretation of average entropy production rate) of size $p\,
\sigma_+$ on one of many segments of duration $\tau$ on a long phase
space trajectory of the dynamical system modeling the shearing f\/luid
in a nonequilibrium stationary state; here $\sigma(x)$ will denote the
phase space contraction rate near a phase point $x$ ({\it i.e.} the
divergence of the equations of motion) and $\sigma_+$ is the average
phase space contraction rate over positive infinite times so that $p$
is a dimensionless characterization of the phase space contraction
(with time average $1$). The approximation within which eq. (\ref{1})
was observed was very convincing \cite{[ECM93]}.

Under suitable assumptions, see below, a more precise formulation of
eq. (\ref{1}) was derived in \cite{[GC95]}:
\begin{equation}
\lim_{\tau \rightarrow \infty} \frac{1}{\tau \sigma_+}
{\rm{ln}} \frac{\pi_\tau(p)}{\pi_\tau(-p)} = p
\label{2}
\end{equation}
where the validity of the f\/luctuation relation eq. (\ref{1}) for
asymptotically long times $\tau$ is more clearly expressed.  Later
several other computer experiments have confirmed the relation
eq. (\ref{2}), \cite{[BGG97]}, \cite{[BCL98]}, \cite{[LLP98]}.

The original computer experiment, \cite{[ECM93]}, was inspired by a
theoretical argument for the relative probabilities to find a phase
space trajectory segment of length $\tau$ in a state $x$ with phase
space contraction rate $p$ and in a state $x'$ with rate $-p$. These
theoretical considerations lead to the correct prediction eq. (\ref{1}),
which was confirmed by the - independently of the theory - carried out
computer experiment.

Evans and Searles \cite{[ES94]} gave a derivation of a theorem which had
a similar form as eq. (\ref{1}).  More precisely: let $E_p$ be the set
of initial conditions of a dynamical system for phase space trajectories
along which the phase space contraction is $e^{-p\sigma_+ T}$ in a time
$T$. We denote by $\mu_L(E_p)$ its Liouville measure.  Similarly, let
$\mu_L(E_{-p})$ be the Liouville measure of the corresponding set of
phase space trajectories along which the phase space contraction in time
$T$ is $e^{p\sigma_+ T}$. Quite generally, and in all models considered
in the literature relevant here, $E_{-p}=I S_TE_p$, if $S_t$ is the time
evolution (Liouville) operator of the system, and if $I$ denotes the
time reversal operation, so that $t \rightarrow S_t x$ is the phase
space trajectory over time $t$ starting at $x$ at $t=0$. Hence $E_{-p}$,
the set of points around which phase space contracts at rate $-p
\sigma_+T$, is obtained by evolving forward over a time $T$ those in
$E_p$ (which would contract by $p \sigma_+T$) and then inverting the
velocities by the time reversal operator $I$.  In fact, the sets $E_p$
and $E_{-p}$ are those considered by Evans and Searles in \cite{[ES94]}.

Then the proof in \cite{[ES94]} is the following:
\begin{eqnarray}
\frac{\mu_L(E_p)}{\mu_L(E_{-p})} & = &
\frac{\mu_L(E_p)}{\mu_L(I S_TE_p)} = \cr
= \frac{\mu_L(E_p)}{\mu_L(S_TE_p)} & = &
\frac{\mu_L(E_p)}{\mu_L(E_p)
e^{-p \sigma_+T}}  
= e^{p \sigma_+T} 
\label{3}
\end{eqnarray}
where one has used that the Liouville distribution is time reversal
invariant, {\it i.e.} $\mu_L(E)\equiv \mu_L(I E)$ (although it is not
stationary) to get the second equality as well as the definition of
phase space contraction in the third equality.

The arbitrary time interval $T$ includes the short times referring to
the transient behavior of the system before reaching the nonequilibrium
stationary state. In the derivation of eq. (\ref{3}) only time reversal
symmetry is used.  Later, \cite{[ES98]}, it was argued that under this
assumption alone, eq. (\ref{3}) also holds in the nonequilibrium
stationary state $\mu_\infty$, since eq. (\ref{3}) is valid for any $T$
and the Liouville distribution $\mu_L$ would evolve in a sufficiently
long time $T$ into a distribution $S_T\mu_L$ arbitrarily close to a
nonequilibrium stationary state $\mu_\infty$.

Therefore the eq. (\ref{3}) was reinterpreted in \cite{[ES98]} and
asserted to be identical to eq. (\ref{2}), which, however, refers to
the statistics of trajectory segments, along a trajectory in a chaotic
nonequilibrium stationary state $\mu_\infty$, {\it not to the statistics of
independent trajectory histories emanating from the initial Liouville
distribution} $\mu_L$ under the time reversibility assumption.

In 1995 the authors proved eq. (\ref{2}) based on a dynamical
assumption, called {\it Chaotic Hypothesis} (CH), which assured strong
chaoticity (``Anosov system-like behavior'') for the systems for which
eq. (\ref{2}) held.  In that work the name Fluctuation Theorem (FT)
was first introduced for eq. (\ref{2}), and was proposed as an
explanation for the experimental result eq. (\ref{1}).  We will call
this the GCFT.

It is worthwhile to emphasize again that, while the right hand sides
of the eqs. (\ref{1}) and (\ref{3}) look very similar, they, as well as
the left hand sides of these equations, really refer to entirely
different physical situations.

The eq. (\ref{3}), \cite{[ES94]}, holds for any $T$ on trajectories
with initial data sampled from the Liouville distribution at $t=0$ and
it can be considered as a simple, but interesting, consequence, for
reversible systems, of the very definition of phase space contraction.
We will call it here the ESI, where the I refers to ``identity''.  The
ESI is much more general than the FT in eq. (\ref{2}), which needs,
{\it in addition} to phase space contraction ($\sigma_+>0$) and time
reversal symmetry, {\it also} the Chaotic Hypothesis. The proof of the
ESI, fully described in eq. (\ref{3}) above, is identical in essence
to the proof in \cite{[ES94]} which is much more involved.

In order to illustrate the fundamental difference between the two 
theorems we first give an example of a case where the more general
ESI eq. (\ref{3}) holds, while the GCFT eq. (\ref{2}) does not.

To that end we consider a single charged particle in a periodic box,
with charge $e$ moving in an electric field ${\bf{E}}$, i.e. a Lorentz
gas without scatterers, and subject to a Gaussian thermostat (to
obtain a nonequilibrium stationary state):
\begin{equation}
{\bf{\dot{q}}} = {\bf{p}}, \qquad
{\bf{\dot{p}}}  =  e {\bf{E}} - \alpha {\bf{p}}
\label{4}
\end{equation}
where the ``thermostat'' force $-\alpha {\bf p}$, with 
$\alpha = \frac{ e {\bf{E}}\cdot{\bf p}}{|\bf p|^2}$, assures the reaching 
of a nonequilibrium stationary state of this system. 

In this case one can solve explicitly the trivial equations of motion
eq. (\ref{4}) and check that the Liouville distribution $\mu_L$ indeed
evolves towards a stationary state $\mu_\infty$, which is
simply a state in which the particle moves with constant speed
parallel to $\bf E$.  The ESI eq. (\ref{3}) will (of course) hold for
the phase space trajectories of this system sampled with the initial
Liouville distribution $\mu_L$, but it will not be a f\/luctuation
theorem, since there are no f\/luctuations. Also, GCFT's eq. (\ref{2})
will not hold for the phase space trajectory segment f\/luctuations of
this system, which is not a contradiction because the system is not
chaotic.

{}From this simple example follows that the two theorems cannot be
equivalent, and the validity of eq. (\ref{3}) cannot imply much, {\it
without extra assumptions}, on the f\/luctuations (absent in this case)
in the stationary state. Note that eq. (\ref{3}) is an identity which
is always valid in the systems considered. The system of eq. (\ref{4})
is therefore a counterexample to the statement, that eq. (\ref{3})
implies eq. (\ref{2}), {\it i.e.} to the statement, \cite{[ES98]},
that ESI implies GCFT.

Second, and more interestingly, one can try to derive the GCFT
eq. (\ref{2}) from the more general ESI eq. (\ref{3}).  One could then
try to proceed as follows.

First one would need to show that on a subsequent trajectory segment
of length $\tau$, {\it after} time $T$, the ratio of the probabilities
of finding a phase space contraction of $+p\sigma_+\tau$ to that of
finding $-p\sigma_+\tau$ over this segment of length $\tau$, would be
given by $e^{p\sigma_+\tau}$.  Here $p\sigma_+\tau$ is any preassigned
value of the phase space contraction.  However, eq. (\ref{3}) gives no
information whatsoever about those points which after a time $T$
evolve into points which in the next $\tau$ units of time show a phase
space contraction $\pm p\sigma_+\tau$. In other words, the ESI does
not contain the detailed information needed to derive the GCFT.

If one adds the Chaotic Hypothesis to the time reversal symmetry
assumptions made about the dynamical system in the ESI, one could use
Sinai's theorem, \cite{[Si72]}, to assert that such a system, starting
from the initial Liouville distribution $\mu_L$, will indeed approach a
chaotic nonequilibrium stationary distribution $\mu_\infty$ supported
(however) on a fractal attractor $A$ with $0$ Liouville measure
$\mu_L(A)=0$.  This is, in this case, the SRB distribution, $\mu_{SRB}$,
of the system, which was used in \cite{[GC95]}.  However, for a proof of
the GCFT, details of the SRB distribution are needed, which contain just
the details considered in \cite{[GC95]}.  That is, one has to make an
appropriate (Markov) partition of the phase space and assign weights to
the cells of increasingly finer partitions leading, to the SRB
distribution. This then allows one to assign appropriate weights to
those regions ({\it of $0$ Liouville measure})  in phase space that
will give rise, to phase space contractions on trajectory segments
$\tau$ of $\pm p\sigma_+\tau$ leading to the GCFT.

A more advanced comparison between the ESI and GCFT requires a more
quantitative statement of the latter result. Namely, \cite{[GC95]},
eq. (\ref{2}) can be derived from the stronger relation:
\begin{equation}
\frac{\pi_\tau(p)}{\pi_\tau(-p)} = e^{(p\,\sigma_++
O(\frac{T_\infty }{\tau}))\,\tau}
\label{5}
\end{equation}
where $T_\infty $ is a time scale of the order of magnitude of the time
necessary in order that the distribution $S_T\mu_L$, into which the
Liouville distribution $\mu_L$ evolves in time $T$, be ``practically''
indistinguishable from the stationary state that we denote by
$\mu_\infty$.  Here the validity of the f\/luctuation relation (\ref{1})
for asymptotically long times $\tau$ is more clearly expressed.  The
esistence of the time $T_\infty $ and its role in bounding the error
term in eq. (\ref{5}) are among the {\it main results of} \cite{[GC95]}.
The time $T_{\infty}$ appears in \cite{[GC95]} as the range of the
potential that generates the representation of $\mu_\infty$ as a Gibbs
state using a symbolic dynamic representation of the SRB distribution on
the Markov partition, \cite{[Si72]}.

Examining the ESI derivation above, one easily sees that the following
\begin{equation}
\frac{S_T\mu_L(E_p)}{S_T\mu_L(E_{-p})}=
\frac{\mu_L(S_{-T}E_p)}{\mu_L(S_{-T}E_{-p})}= 
e^{(p\,\sigma_++ O(\frac{T}\tau))\,\tau}
\label{6}
\end{equation}
can also be derived, \cite{[Le]}; it is important to note that the
argument leading to eq. (\ref{3}) {\it cannot say more than this}: in
particular one must justify why $T$, which in principle should be large,
{\it strictly speaking infinite}, so that $S_T\mu_L$ be identifiable
with $\mu_{SRB}$, {\it can in fact be taken smaller than
$\tau$}. Justifying this requires assumptions (as the above
counterexample indicates) like the mentioned Gibbs property of the SRB
distribution which is even stronger than requiring exponential decay of
correlations (also implied by the CH). Otherwise, without extra
assumptions, one has to say that $T$ has to be taken infinite first with
the result that eq. (\ref{6}), hence eq. (\ref{3}), {\it becomes empty}
in content.  Of course one can argue that ``on physical grounds'' $T$
needs not to be taken infinite but just as large as some characteristic
time scale for the approach to the attractor: but the precise meaning of this,
and the assumptions under which it can be stated, is precisely what
needs to be determined particularly because in nonequilibrium systems
the attractor $A$ is {\it fractal} with $\mu_L(A)=0$ and one can very
well doubt that the $S_T\mu_L$ distribution is {\it ever close enough to
the $\mu_\infty$ distribution to allow comparing} eq. (\ref{5}) and
(\ref{6}). This requires a convincing argument, were it only because
$S_T\mu_L(A)=0$.

\vspace{7mm}
 
\noindent {\small{\it Acknowledgements:} E.G.D.C. is very much
indebted to L. Rondoni for many very clarifying discussions. The authors
are also indebted to C. Liverani for suggestions on the proof of 
eq. (\ref{3}) and to F. Bonetto, J. L. Lebowitz, and L. Rondoni for
helpful discussions. E.G.D.C. gratefully acknowledges support from DOE
grant DE-FG02-88-ER13847, while G.G. that from Rutgers University's 
CNR-GNFM, MPI.}

\def\revtex{R\raise2pt\hbox{E}VT\lower2pt\hbox{E}X}
\revtex
\end{document}